\documentclass[12pt]{iopart}
\usepackage{times}
\usepackage{graphicx}

\newcommand{\nix}[1]{}
\usepackage{bm}

\begin{document}

\title{ Monopolar Optical Orientation of Electronic Spins in Semiconductors}
\author{S.A.~Tarasenko$^1$, E.L.~Ivchenko$^1$, V.V.~Bel'kov$^1$,
S.D.~Ganichev$^{2}$, D.~Schowalter$^{2}$,  Petra~Schneider$^2$, M.~Sollinger$^2$,
W.~Prettl$^2$, \\V.M.~Ustinov$^1$,  A.E.~Zhukov$^1$,  and L.~E.~Vorobjev$^3$}
\address{
$^1$ A.F.~Ioffe Physico-Technical Institute, 194021 St. Petersburg,
Russia
\\$^2$ Fakult\"{a}t f\"{u}r Physik, Universit\"{a}t Regensburg, D-93040 Regensburg,
Germany\\
$^3$ St. Petersburg State Technical University, 195251 St.~Petersburg, Russia
}

\begin{abstract}
It is shown that absorption of circularly polarized infrared radiation
due to intraband (Drude-like) transitions
in $n$-type bulk semiconductors and
due to intra-subband  or inter-subband  transitions in quantum well (QW)
structures results
in a monopolar spin orientation of free electrons. Spin
polarization in zinc-blende-structure based QWs
is demonstrated  by the observation of the  spin-galvanic  and
the circular photogalvanic effects. The monopolar  spin orientation
in $n$-type materials is shown to be possible if an admixture of
valence band states to the conduction band wave function and the
spin-orbit splitting of the valence band are taken into account.
\end{abstract}

\section{Introduction}

Absorption of circularly polarized light in semiconductors may
result in spin polarization of photoexcited carriers. This
phenomenon of optical orientation is well known for interband
transitions in  semiconductors~\cite{Meier}.
At interband excitation with circularly polarized light
transitions from the valence to the conduction band are
allowed only if the angular momentum
is changed by $\pm 1$. These selection
rules lead to the spin orientation of carriers with the sign and degree of
polarization depending on the light helicity. While optical
orientation at interband excitation has been
widely studied, it is not obvious that the absorption of infrared
radiation  due to
intraband optical transitions can result in a spin polarization.

In this paper we show that free carrier absorption of circularly polarized
radiation due to both, indirect optical transitions for bulk semiconductors
or quantum well structures, and direct transitions between size-quantized
subbands also leads to spin orientation of free electrons. This optical
orientation has not been considered previously and may be referred to as
`monopolar spin orientation' because only one type of carriers is excited,
electrons in  $n$-type
materials, holes in $p$-type materials.
We present theoretical and experimental
results on the monopolar optical orientation of intraband absorption by
free carriers in $n$-type zinc-blende-structure bulk semiconductors and low
dimensional structures. We show that the monopolar optical orientation of
electrons can be obtained if an admixture of the $\Gamma_7$ and $\Gamma_8$
valence band states to the conduction band wave functions is taken into
account.
   \begin{figure}
   \begin{center}
   \begin{tabular}{c}
    \mbox{\includegraphics[height=3.6cm]{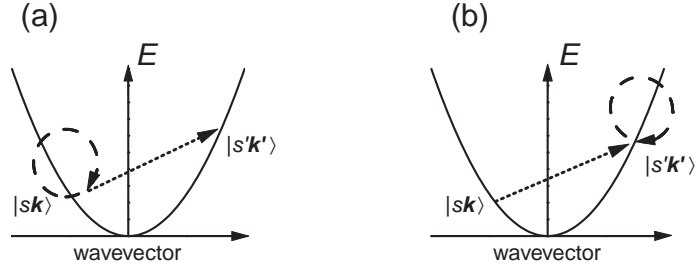}}
   \end{tabular}
   \end{center}
   \caption[example]
   { \label{band1}
Schematic representation of indirect intraband
optical transitions with intermediate states in the same band. Dashed
and dotted curves indicate the electron-photon interaction and the electron
momentum scattering. Figures~(a) and (b) correspond to the first and the
second term in Eq.~(\protect \ref{matrelem}).
 }
   \end{figure}

\section{Monopolar optical spin orientation in bulk semiconductors}

Absorption of infrared light by free electrons in $n$-type semiconductors
(Drude-like absorption)  occurs by indirect intraband  transitions
where momentum conservation law is satisfied due to acoustic phonons,
optical phonons, static defects etc. The  intraband
absorption of the light involving both
electron-photon interaction and scattering is described by
second-order processes with virtual intermediate states.
The compound matrix elements for such kind of transitions
with the initial and final states in the same band has the standard form
\begin{equation}\label{matrelem}
M_{cs'\bm{k}' \leftarrow cs\bm{k}}=\sum_{\nu} \left( \frac{
V_{cs'\bm{k}',\,\nu\bm{k}}\,R_{\,\nu\bm{k},cs\bm{k}} }
{E_{\nu\bm{k}}-E_{c\bm{k}}-\hbar\omega} + \frac{
R_{\,cs'\bm{k}',\,\nu\bm{k}'}\,V_{\nu\bm{k}',cs\bm{k}} }
{E_{\nu\bm{k}'}-E_{c\bm{k}} \pm \hbar\Omega_{\bm{k}-\bm{k}'}}
\right) \;.
\end{equation}
Here $E_{c\bm{k}}$, $E_{c\bm{k}'}$ and $E_{\nu}$ are the electron
energies in the initial $|c,s, \bm{k} \rangle$, final $|c,s',
\bm{k}' \rangle$ and intermediate $|\nu \rangle$ states, respectively,
$s$ is the
spin index, $\bm{k}$ is the electron wavevector, $R$ is the
matrix element of electron interaction with the electromagnetic wave,
$V$ is the matrix element of electron-phonon or electron-defect interaction,
and $\hbar\Omega_{\bm{k}-\bm{k}'}$ is the
energy of the involved phonon. At static defect assisted
scattering $\hbar\Omega =0$. The sign $\pm$ in Eq.~(\ref{matrelem})
correspond to  emission and absorption of phonons.
A dominant contribution to the light absorption due to indirect transitions in
the conduction band is caused by processes with intermediate states in the
same band (see  Fig.~1).  The corresponding matrix element has the form
\begin{equation}\label{Mabs}
M_{cs' \bm{k}' \leftarrow cs \bm{k}} \propto
\frac{V_{cs' \bm{k}',cs\bm{k}}}{\hbar \omega} \, \frac{m_0}{\hbar}
\, \bm{e} \cdot \left( \frac{\partial E_{c \bm{k}'}}{\partial  \bm{k}'}
- \frac{\partial E_{c \bm{k}}}{\partial  \bm{k}} \right) \:,
\end{equation}
where $m_0$ is the free electron mass, and $\bm{e}$ is
the unit vector of the electric field polarization.
The absolute value of the matrix
element in Eq.~(\ref{Mabs}) is independent of the degree of
circular polarization of radiation, $P_{circ}$.
Hence, the intraband transitions with  intermediate
states in the same band {\it do not}  contribute to the optical
orientation.
   \begin{figure}
   \begin{center}
   \begin{tabular}{c}
\mbox{\includegraphics[height=5.06cm]{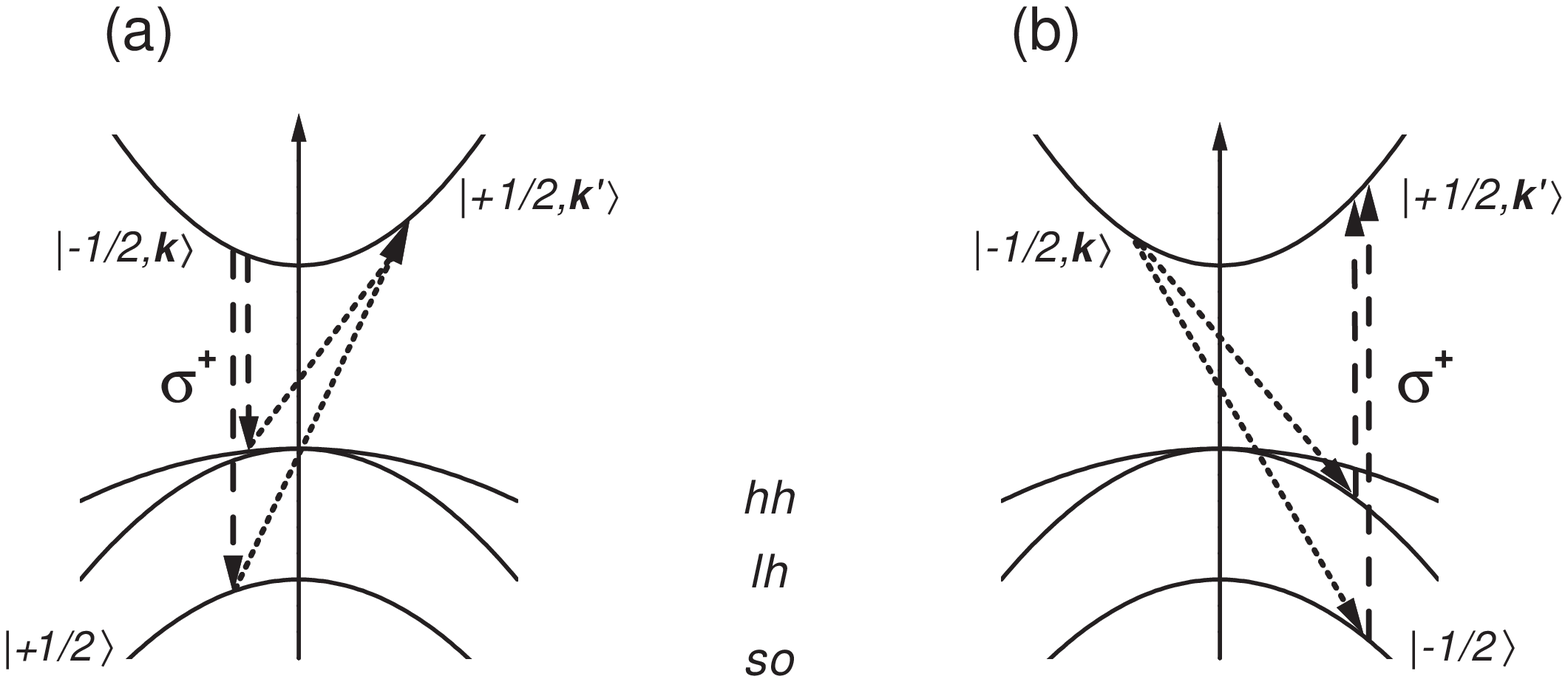}}
   \end{tabular}
   \end{center}
   \caption[example]
   { \label{band2}
Sketch of indirect intraband optical transitions
with intermediate states in the valence band. Dashed
and dotted curves indicate the electron-photon interaction and the electron
momentum scattering.
 }
   \end{figure}

Spin orientation caused by intraband absorption of circularly
polarized light can be obtained
considering processes with intermediate states in the
valence band and taking into account its spin-orbit
splitting. Fig.~2 demonstrates schematically the
spin orientation at intraband absorption of
right handed circularly polarized
light ($\sigma^{+}$).
Because of the selection
rules for interband optical matrix elements, the electron transitions
with  spin reversal $-1/2 \rightarrow +1/2$ are possible via
intermediate states in the light-hole and spin-orbit split subbands,
while the opposite processes, $+1/2 \rightarrow -1/2$ are forbidden.
We assume the photon energy $\hbar\omega$ to exceed the typical
energy of equilibrium electrons, $\hbar\omega \gg k_B T$ for
a nondegenerate electron gas at the temperature $T$, or $\hbar\omega \gg
\varepsilon_F$ for the case of degenerate statistics with the
Fermi energy $\varepsilon_F$. Considering acoustic-phonon-assisted
light absorption as an example, the following
equation for the spin generation rate is obtained
\begin{equation}\label{spinbulk}
\dot{\bm{S}} = \frac16 \frac{\Xi^2_{cv}}{\Xi^2_c}
\frac{\Delta^2_{so}}{E_g(E_g+\Delta_{so})(3E_g+2\Delta_{so})} \, K
\, \bm{\hat{e}} \, I \, P_{circ} \:.
\end{equation}
Here $\Xi_{cv}$ and $\Xi_c$ denote the interband and intraband
deformation potential constants, respectively,
$E_g$   stands for
the energies of the band gap and $\Delta_{so}$ for
the valence band spin-orbit
splitting, $I$ is the light intensity,
and $\bm{\hat{e}}$
is the unit vector in the light propagation direction. The
factor $K$ in Eq.~(\ref{spinbulk}) is the coefficient of the light
absorption under phonon-assisted intraband optical transitions.
It is determined by the dominant processes with intermediate
states in the same  band, and has the form
\begin{equation}
K=\frac{4 \alpha}{3 n_{\omega}}
\left(\frac{\Xi_c}{\hbar\omega}\right)^2 \frac{k_B T}{\rho v_s^2}
\left(\frac{2m^* \omega}{\hbar}\right)^{1/2} N_e   \:,
\end{equation}
where $\alpha$ and $n_{\omega}$ are the fine structure constant
 and the refraction index of the medium, respectively, $\rho$
is the mass density of the crystal, $v_s$ is the sound
velocity, $m^*$ is the
effective mass of the conduction electron, and $N_e$ is the
carrier concentration.

It should be emphasized that the spin generation depends strongly on the
energy of spin-orbit splitting of the valence band, $\Delta_{so}$.
It is due to the fact that the contributions to the matrix element
giving rise to the optical orientation from the valence band
$\Gamma_8$ and spin-orbit split band $\Gamma_7$ have the opposite
signs. An estimation for GaAs material shows that the typical value
of spin generated per one absorbed photon is of order $10^{-6}$
at a photon energy of $\hbar \omega = 10$ meV.\\

\section{Monopolar optical spin orientation in QWs}

\subsection{Experimental technique}
Experimentally, the monopolar spin orientation has been investigated
on MBE (001)-grown
$n$-GaAs/AlGaAs QW of width $d_W= 7$~nm
and $n$-In$_{0.2}$Ga$_{0.8}$As
QWs of 7.6~nm width.
Samples with free carrier
densities of about $2\cdot 10^{11}$ cm$^{-2}$ were studied in the
temperature range
from liquid helium to room temperature. A pair of ohmic contacts was
centered on opposite sample edges along the direction $x \parallel
[1\bar{1}0]$ (see in Fig.~3).

A high power pulsed mid-infrared (MIR) TEA-CO$_2$ laser and a far-infrared
(FIR) NH$_3$ laser were  used as radiation sources delivering 100\,ns
pulses with radiation power $P$ up to 100\,kW. Several lines of the CO$_2$
laser between 9.2\,$\mu$m and 10.6\,$\mu$m and of the NH$_3$-laser~\cite{PhysicaB99}
between $\lambda$ = 76\,$\mu$m and 280\,$\mu$m were chosen  for
excitation in the MIR and FIR range, respectively.
Excitation of samples  by FIR radiation with photon energy less than the
separation of size-quantized subbands leads to absorption caused by
indirect optical transitions
in the lowest subband (Drude absorption). The
MIR radiation induces direct optical transitions between the first and the
second subband of QWs.

The laser light polarization was modified from
linear to circular using a Fresnel rhombus and  quartz $\lambda/4$ plates
for MIR and
FIR radiation, respectively. The
helicity of the incident light was varied
according to $P_{circ} = \sin{2
\varphi}$ where $\varphi$ is the angle between the initial
plane of linear polarization and the optical axis of the
polarizer. Spin polarization has been investigated making use of the
circular photogalvanic effect (CPGE)~\cite{PRL01} and the spin-galvanic
effect (SGE)~\cite{Nature02}. The experimental procedure is sketched in Fig~3.
For investigation of the spin-galvanic effect
an in-plane magnetic
field $B$ up to 1~T has been applied.
The current $j$ generated by polarized light in the unbiased structures
was measured via the voltage drop across a 50~$\Omega$ load resistor in a
closed circuit configuration. The voltage was recorded with a storage
oscilloscope. The measured current pulses of 100~ns
duration reproduce the temporal structure of the laser pulses.

    \begin{figure}
   \begin{center}
   \begin{tabular}{c}
    \mbox{\includegraphics[height=5.6cm]{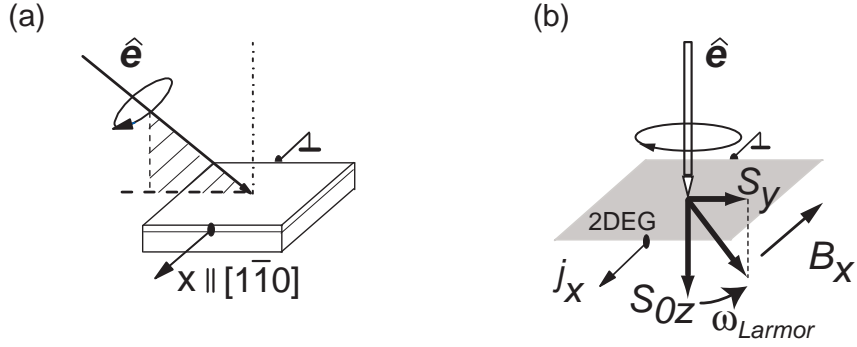}}
   \end{tabular}
   \end{center}
   \caption[example]
   { \label{PRL01}
Experimental set-up used for measuring of (a) the circular photogalvanic
effect and (b) the spin-galvanic effect in (001)-grown QWs. The
photocurrent occurs in the first case at oblique incidence only and in the
second case at normal incidence in combination with an in-plane magnetic
field $B_x$. The current flow in both, the circular photogalvanic effect
and the spin-galvanic effect, is driven by an asymmetric distribution of
carriers in {\boldmath$k$}-space in systems with lifted spin degeneracy due
to {\boldmath$k$}-linear terms in the Hamiltonian. The
spin-galvanic effect is caused by asymmetric spin-flip scattering of spin
polarized carriers and it is determined by the process of spin relaxation.
If spin relaxation is absent, the effect vanishes. In contrast, the
circular photogalvanic effect is the result of selective photoexcitation of
carriers in {\boldmath$k$}-space with circularly polarized light due to
optical selection rules.

 }
   \end{figure}

\subsection{Intra-subband transitions in QWs}
In quantum well structures absorption of infrared radiation may be achieved
by indirect intra-subband optical transitions and, for photon energies
being in resonance with the energy distance between size quantized
subbands, by direct transitions between these subbands. To obtain
absorption caused by indirect transitions we used FIR radiation in the
range from 76~$\mu$m to 280~$\mu$m (corresponding photon energies are from
16\,meV to 4.4\,meV). The experiments were carried out on GaAs and InAs QWs
with the energy separation $\Delta E=E_2-E_1$ between $e1$ and $e2$
size-quantized subbands equal to 120~meV and 112~meV, respectively.
Therefore the photon energies used are much smaller than $\Delta E$ and
absorption is caused by indirect intra-subband optical transitions. With
illumination of (001)-grown GaAs and InAs QWs at oblique incidence of FIR
radiation a current signal proportional to the helicity
$P_{circ}$ has been observed (see Fig.~4a) indicating the circular
photogalvanic effect~\cite{PRL01}. At normal incidence of radiation, where
the CPGE vanishes, the spin-galvanic current~\cite{Nature02} is also
observed applying an in-plane magnetic field (see Fig.~4b). Both effects
are due to spin orientation. Therefore the observation of the CPGE and the
spin-galvanic effect gives clear evidence that the absorption of far
infrared circularly polarized radiation results in spin orientation. We
note, that monopolar orientation has also been observed for $p$-type QW
structures, but this is out of the scope of the present paper.

   \begin{figure}
   \begin{center}
   \begin{tabular}{c}
    \mbox{\includegraphics[height=5.6cm]{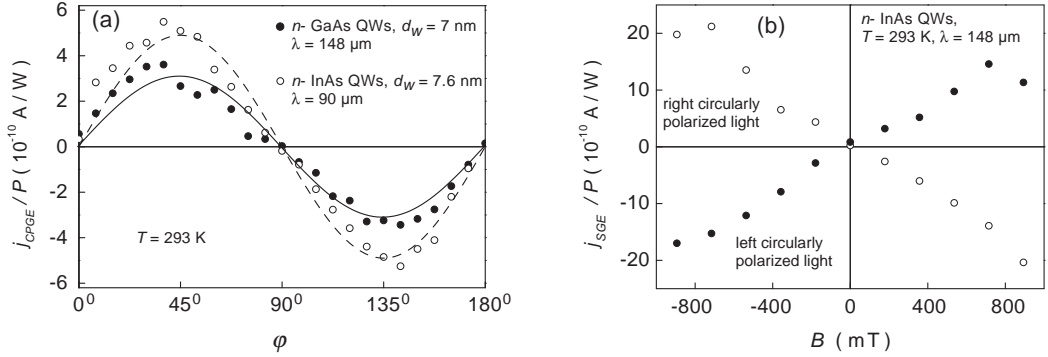}}
   \end{tabular}
   \end{center}
   \caption[example]
   { \label{PRL01}
Monopolar spin orientation due to indirect intra-subband transitions within
the $e1$ conduction subband in QW structures demonstrated by (a) the circular
photogalvanic effect: the photocurrent in QWs normalized by light power~$P$ is
plotted as a function of the phase angle $\varphi$ defining helicity. Full
lines are fitted using one parameter according to $j \propto \sin 2
\varphi$, (b) the spin-galvanic effect: the figure shows magnetic field
dependence of the photocurrent normalized by light power~$P$. Measurements
are presented for $T=~$293~K. }
\end{figure}

Monopolar optical orientation caused by indirect intra-subband transitions
in QWs can be obtained in the similar way as that for bulk semiconductors
(see previous section). For this particular mechanism of the monopolar optical
orientation, the spin generation rate has been derived for two scattering
processes: acoustic phonons and elastic scattering on static defects.

Considering acoustic-phonon-assisted processes with virtual intermediate
states in the heavy-hole $hh$, light-hole $lh$, and spin-orbit split
$so$ subband, we obtain for the spin generation rate
\begin{equation}\label{spin2D}
\dot{\bm{S}}_{ph} = \frac16 \frac{\Xi^2_{cv}}{\Xi^2_c}
\frac{\Delta^2_{so}}{E_g(E_g+\Delta_{so})(3E_g+2\Delta_{so})}
\left[ \bm{\hat{e}}_{\parallel} + \frac{d_W}{3} \sqrt{\frac{2 m^* \omega}
{\hbar}} \, \bm{\hat{e}}_{z} \right] \eta_{ph} \, I \, P_{circ} \:,\,\,\,\,
\end{equation}
where $d_W$ is the width of a QW, $\eta_{ph}$ is the fraction of the energy
flux absorbed in a QW under normal incidence. It is given by
\begin{equation}
\eta_{ph}=\frac{3 \pi \alpha}{n_{\omega}}
\left(\frac{\Xi_c}{\hbar\omega}\right)^2
\frac{k_B T}{\rho a v_s^2} \, N_e   \:,
\end{equation}
where $N_e$ is the concentration of the two-dimensional electron gas.

In the case of elastic scattering on
static impurities, the expression for spin generation rate has
the form
\begin{equation}\label{spin2Ddef}
\dot{\bm{S}}_{imp} = \frac23
\frac{\Delta^2_{so}}{E_g(E_g+\Delta_{so})(3E_g+2\Delta_{so})}
\left[ \, \frac{V^2_{\parallel}}{V^2_{0}} \, \bm{\hat{e}}_{\parallel} +
\frac{V^2_{z}}{V^2_{0}} \, \bm{\hat{e}}_{z} \right] \eta_{imp} \, I \, P_{circ} \:.
\end{equation}
Here $V_0$ is the intraband matrix element of scattering, the factors
$V_{z}$ and $V_{\parallel}$ describe impurity-induced mixing of the
conduction band Bloch function $S$ with the valence band Bloch functions $Z$
and $X,Y$ respectively. The fraction of the energy flux absorbed in the
QW at impurity-assisted intrasubband optical transitions under normal incidence
is given by
\begin{equation}
\eta_{imp}=\frac{2 \pi \alpha}{n_{\omega}}
\left(\frac{V_0}{\hbar\omega}\right)^2 N_e \, N_d  \:,
\end{equation}
where $N_d$ is the concentration of scatterers.

\subsection{Direct inter-subband transitions in QWs}

Absorption of radiation in the range from 9~$\mu$m to 11~$\mu$m in our GaAs
and InAs samples is dominated by resonant direct inter-subband optical
transitions between the first and the second subband. Fig.~5b shows
the resonance behaviour of absorption measured in GaAs QWs by making use of
transmission Fourier spectroscopy in a
multiple-reflexion
waveguide geometry (see inset Fig.~5b). Applying MIR
radiation of the CO$_2$ laser, which causes direct transitions in GaAs and
InAs QWs, the circular photogalvanic current at oblique incidence (Fig.~5a)
and the spin-galvanic current at normal incidence of radiation (Fig.~5b)
have also been observed. The wavelength dependence of the spin-galvanic
effect obtained between 9.2~$\mu$m and 10.6~$\mu$m repeats the spectral
behaviour of direct inter-subband absorption. This unambiguously
demonstrates that in this case the spin orientation of $n$-type QWs is
obtained by inter-subband transitions.

 \begin{figure}
   \begin{center}
   \begin{tabular}{c}
    \mbox{\includegraphics[height=5.6cm]{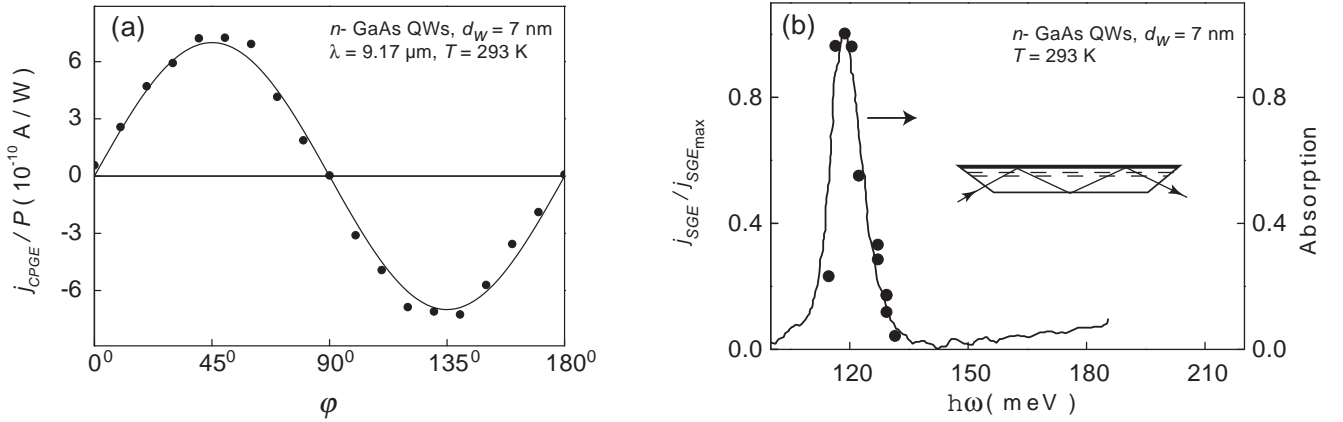}}
   \end{tabular}
   \end{center}
   \caption[example]
   { \label{SG}
Monopolar spin orientation due to direct inter-subband transitions between
$e1$ and $e2$ conduction subbands in QW structures demonstrated by (a)
the photogalvanic effect: the photocurrent in QWs normalized by light power~$P$ is
plotted as a function of the phase angle $\varphi$ defining helicity. Full
lines are fitted using one parameter according to $j \propto \sin 2
\varphi$, (b) the spin-galvanic effect: the figure shows the
spectral dependence
of the magnetic field induced photocurrent (dots). Data are presented for
optical excitation at normal incidence of right-handed circularly polarized
radiation. A magnetic field of $B= 1$~T was used. For comparison the
absorption spectrum is shown by the full line. The absorption has been
determined by transmission measurements making use of a multiple-reflexion
waveguide geometry shown in the inset. Results are plotted for (001)-grown
GaAs QWs of 7~nm width at room temperature.}

   \end{figure}

We would like to emphasize that
spin sensitive inter-subband transitions in
$n$-type QWs have been observed
at normal incidence when there is no component of the electric
field of the radiation normal to the plane of the QWs.
Generally it is
believed that inter-subband transitions in $n$-type QWs
can only be excited by infrared
light polarized in the growth direction $z$ of the QWs~\cite{book}.
Furthermore such transitions are spin insensitive and, hence, do not lead to
optical orientation. Since the argument, leading to these selection rules, is
based on the effective mass approximation in a single band model, the
selection rules are not rigorous.

In order to explain the observed spin orientation as well as the absorption
of light polarized in the plane of the QW we show that
a ${\bf k}\cdot{\bf p}$ admixture of valence band states to the
conduction band wave functions has to be taken into account.
Calculations yield that inter-subband absorption of circularly polarized light propagating
along $z$  induces only spin-flip transitions resulting in  100\% optical orientation of
photoexcited carriers. In this geometry the  spin generation rate for resonant inter-subband transitions has the form
\begin{equation}
\dot{S}_z = \frac{128 \alpha}{9 n_{\omega}} \,
\frac{\Delta^2_{so}(2E_g+\Delta_{so})^2 E_1}
{E^2_g(E_g+\Delta_{so})^2(3E_g+2\Delta_{so})^2} \, \frac{\hbar^2 N_e}{m^* }
\,\delta(\hbar\omega - E_1 + E_2) \,I \, P_{circ} \: \,\,\,\,\,
\end{equation}
where the function $\delta$ describes the resonant behaviour
of the inter-subband transitions.

\section{Summary}

In conclusion, our results demonstrate that in $n$-type bulk semiconductors
as well as QW structures monopolar spin orientation can be achieved
applying circularly polarized
radiation with photon energies less than the fundamental energy
gap. Spin orientation has been observed by indirect intra-subband
absorption as well as by inter-subband transitions. It is shown that
monopolar spin orientation in $n$-type materials becomes possible if an
admixture of valence band states to the conduction band wave function and
the spin-orbit splitting of the valence band are taken into account. We
emphasize that the spin generation rate under monopolar optical orientation
depends strongly on the energy of spin-orbit splitting of the valence band,
$\Delta_{so}$. It is due to the fact that the valence band $\Gamma_8$ and
the spin-orbit split band $\Gamma_7$ contribute to the matrix element of
spin-flip transitions with opposite signs.

\section{Acknowledgments}
We thank D. Weiss and W. Wegscheider for many helpful discussions.
Financial support from the DFG, the RFFI, the Russian Ministry of Science and
the NATO linkage program is gratefully acknowledged.


\section*{References}

\begin{thebibliography}{20}

\bibitem{Meier}{\it Optical Orientation}, F.\,Meier and
B.P.~Zakharchenya, Eds. (Elsevier, Amsterdam 1984).
%
\bibitem{PhysicaB99} S.D.~Ganichev, Physica B~{\bf 273-274}, 737 (1999).
%
\bibitem{PRL01} S.D.\,Ganichev,
E.L.\,Ivchenko, S.N.\,Danilov,  J.\,Eroms, W.\,Wegscheider, D.\,Weiss, and W.\,Prettl,
Phys.\,Rev.\,Lett. {\bf 86}, 4358 (2001).
%
\bibitem{Nature02} S.D.~Ganichev, E.L.~Ivchenko,
V.V.~Bel'kov, S.A.~Tarasenko, M.~Sollinger, D.~Weiss, W.~Wegscheider, and
W. Prettl,
{\it Nature} (London) {\bf 417}, 153 (2002).

\bibitem{book} E.L.~Ivchenko, and  G.E.~Pikus, {\it Superlattices and Other
Heterostructures. Symmetry
and Optical Phenomena},
(Springer, Berlin 1997).

\end{thebibliography}
\end{document}